\documentclass[]{spie}  %>>> use for US letter paper
\usepackage{amsmath,amsfonts,amssymb}
\usepackage{graphicx}
\usepackage{upgreek}
\usepackage[colorlinks=true, allcolors=blue]{hyperref}
\usepackage{svg}

\title{The iLocater cryostat and thermal control system: enabling extremely precise radial velocity measurements for diffraction-limited spectrographs}

\author[a]{Jonathan Crass}
\author[a,b]{Nandini Sadagopan}
\author[b]{Matthew Misch}
\author[c]{Alexa Rizika}
\author[d]{Brian Sands}
\author[d]{Matthew Engstrom}
\author[a]{Justin R. Crepp}
\author[d]{James Smous}
\author[a]{Jeffrey Chilcote}
\author[e]{Louis G. Fantano}
\author[a]{Michael VanSickle}
\author[f]{Frederick R. Hearty}
\author[g]{Matthew J. Nelson}
\affil[a]{Department of Physics \& Astronomy, 225 Nieuwland Science Hall, Notre Dame, IN 46556, USA}
\affil[b]{Department of Aerospace \& Mechanical Engineering, 365 Fitzpatrick Hall of Engineering, Notre Dame, IN 46556, USA}
\affil[c]{Department of Civil \& Environmental Engineering \& Earth Sciences, 156 Fitzpatrick Hall of Engineering, Notre Dame, IN 46556 USA}
\affil[d]{Engineering \& Design Core Facility, 626 Flanner Hall, Notre Dame, IN 46556 USA}
\affil[e]{NASA Goddard Space Flight Center, 8800 Greenbelt Rd, Greenbelt, MD 20771, USA}
\affil[f]{405 Davey Laboratory, Pennsylvania State University, University Park, PA 16802, USA}
\affil[g]{Department of Astronomy, University of Virginia, Charlottesville, VA 22904-4325, USA}

\authorinfo{Send correspondence to Jonathan Crass: j.crass@nd.edu}

\begin{document} 
\maketitle

\begin{abstract}
Extremely precise radial velocity (EPRV) measurements are critical for characterizing nearby terrestrial worlds. EPRV instrument precisions of $\sigma_{\mathrm{RV}} = 1-10\,\mathrm{cm/s}$  are required to study Earth-analog systems, imposing stringent, sub-mK, thermo-mechanical stability requirements on Doppler spectrograph designs. iLocater is a new, high-resolution ($R=190,500$ median) near infrared (NIR) EPRV spectrograph under construction for the dual 8.4\,m diameter Large Binocular Telescope (LBT). The instrument is one of the first to operate in the diffraction-limited regime enabled by the use of adaptive optics and single-mode fibers. This facilitates affordable optomechanical fabrication of the spectrograph using intrinsically stable materials.

We present the final design and performance of the iLocater cryostat and thermal control system which houses the instrument spectrograph. The spectrograph is situated inside an actively temperature-controlled radiation shield mounted inside a multi-layer-insulation (MLI) lined vacuum chamber. The radiation shield provides sub-mK thermal stability, building on the existing heritage of the Habitable-zone Planet Finder (HPF) and NEID instruments. The instrument operating temperature ($T=80-100\,\mathrm{K}$) is driven by the requirement to minimize detector background and instantaneous coefficient of thermal expansion (CTE) of the materials used for spectrograph fabrication. This combination allows for a reduced thermomechanical impact on measurement precision, improving the scientific capabilities of the instrument.
\end{abstract}


\section{Introduction \& Motivation}
\label{sec:intro}  % \label{} allows reference to this section

Extremely precise radial velocity (EPRV) measurements continue to be critical in detecting and characterizing the properties of nearby exoplanet systems. Recent strategic reports, including the NAS Exoplanet Survey Strategy, Astro2020 Decadal Survey, and NASA/NSF EPRV Working Group Report, have all highlighted the need to continue to advance EPRV capabilities to enable the detection and characterization of Earth-like planets around Sun-like stars \cite{NAP25187, NAP26141, 2021arXiv210714291C}. Instrument precisions at the $\sigma_{\mathrm{RV}} = 1-10\,\mathrm{cm/s}$ level are required to achieve this goal, imposing stringent, sub-mK thermo-mechanical stability requirements on Doppler spectrographs. Even for current generation instruments, providing an optimized and stable environment for a spectrograph reduces instrument systematics, and helps to push the capabilities of existing technologies.

iLocater is a new high-resolution ($R=190,500$ median) near infrared (NIR) EPRV instrument under construction for the dual 8.4\,m diameter Large Binocular Telescope (LBT) \cite{2016SPIE.9908E..19C}. The instrument uses single-mode fibers for illumination, enabled through the use of adaptive optics (AO) for injection, and is one of the first EPRV instruments to operate in the diffraction-limited regime. Operating at the diffraction-limit enables a high-resolution spectrograph to be delivered in a compact instrument volume and allows intrinsically stable, low coefficient of thermal expansion (CTE), materials to be used for optomechanical fabrication \cite{2014Sci...346..809C}. When housed within a thermally stabilized environment, this enables the spectrograph to achieve optimal performance and its science goals.

We present the final design of the iLocater cryostat system which houses the instrument spectrograph. Section \ref{sec:drivers} discusses specific design drivers for the systems and considerations for optimization, followed by an overview of the cryostat system being presented in Section \ref{sec:design}. Section \ref{sec:thermal} presents the cryostat thermal system, its simulated performance and initial results of system cooling capacity. A discussion of system performance and program status is presented in Section \ref{sec:conclusions}.

\section{System Design Drivers \& Optimization}
\label{sec:drivers}  % \label{} allows reference to this section

To achieve the science requirements of the iLocater spectrograph, the cryostat has several constraints on its design and performance. These include a combination of operational and performance constraints, and practical considerations for operation within an observatory environment:

\begin{itemize}
    \item The system operating temperature should be within the regime where thermal background recorded on the instrument detector (H4RG-10) should not impact scientific performance. Additionally, the temperature should be selected to minimize the instantaneous CTE of materials used within the optomechanical system (Section \ref{sec:temperature}).
    \item The system must achieve milli-Kelvin levels of thermal stability over extended periods to provide a baseline for EPRV studies.
    \item The cryostat must impart minimal vibrations to the internal spectrograph system.
    \item The timescale for instrument cooldown and warm-up cycles must not be a burden on general operation. This requires a pumping and cooldown cycle not to exceed 7 days as a goal, with 14 days as a requirement.
    \item There should be minimal routine support required at the observatory for the instrument to operate. This dictates the use of cryocoolers for cooling rather than liquid cryogens.
    \item The location where the cryostat will be installed at the telescope has size and weight limitations that must be adhered to by the completed system.
\end{itemize}

\section{Cryostat Design Overview}
\label{sec:design}  % \label{} allows reference to this section

\begin{figure} [ht]
   \begin{center}
   \begin{tabular}{c} %% tabular useful for creating an array of images 
   \includegraphics[width=\textwidth]{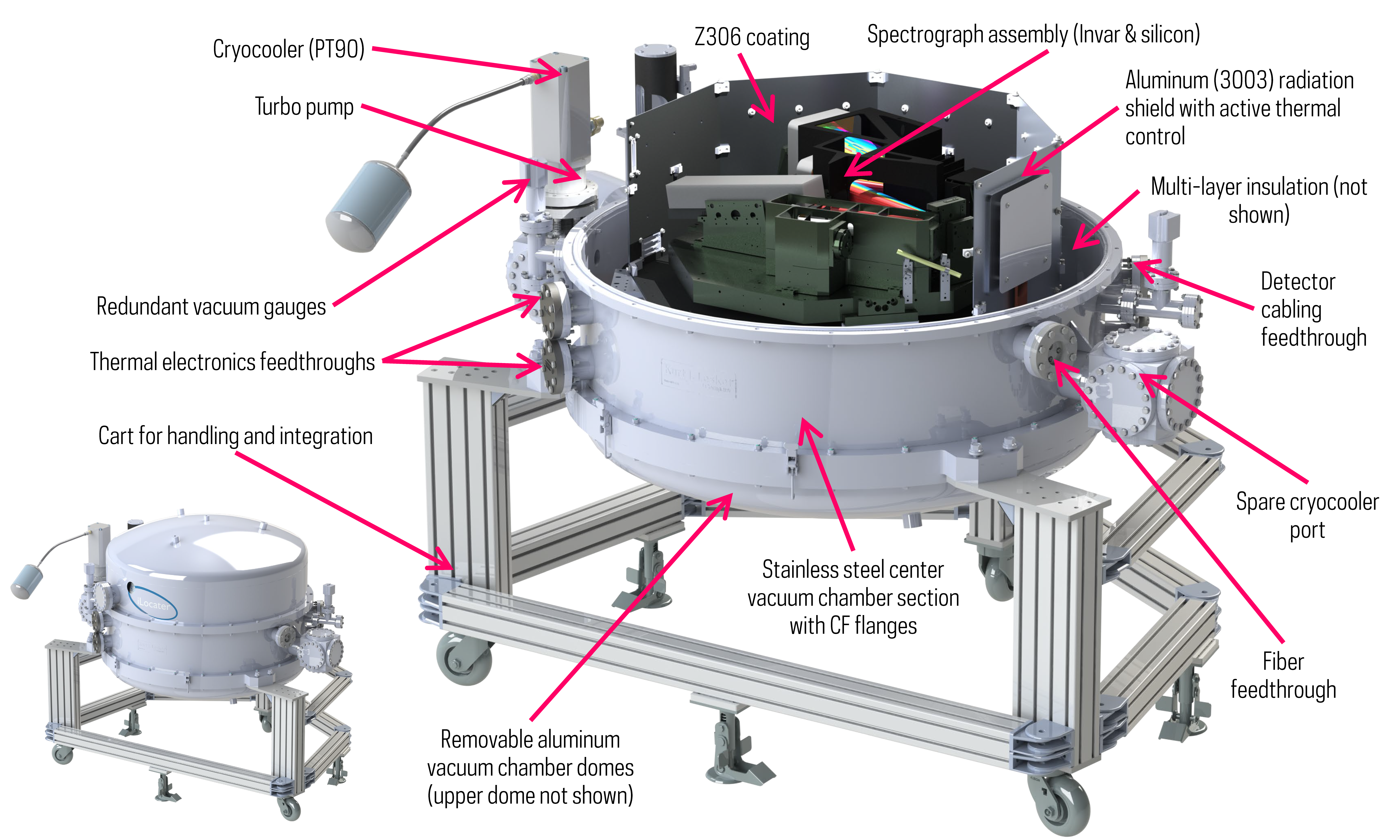}
   \end{tabular}
   \end{center}
   \caption[CAD rendering of the cryostat] 
%>>>> use \label inside caption to get Fig. number with \ref{}
   { \label{fig:cryostat} 
Annotated CAD rendering of the iLocater cryostat showing the instrument spectrograph at the center. The spectrograph is housed inside a thermally stabilized radiation shield which surrounds the spectrograph. The inset figure shows the closed vacuum chamber.}
   \end{figure}

\begin{figure} [ht]
   \begin{center}
   \begin{tabular}{c} %% tabular useful for creating an array of images 
   \includegraphics[width=\textwidth,trim={0 0 0 8cm},clip]{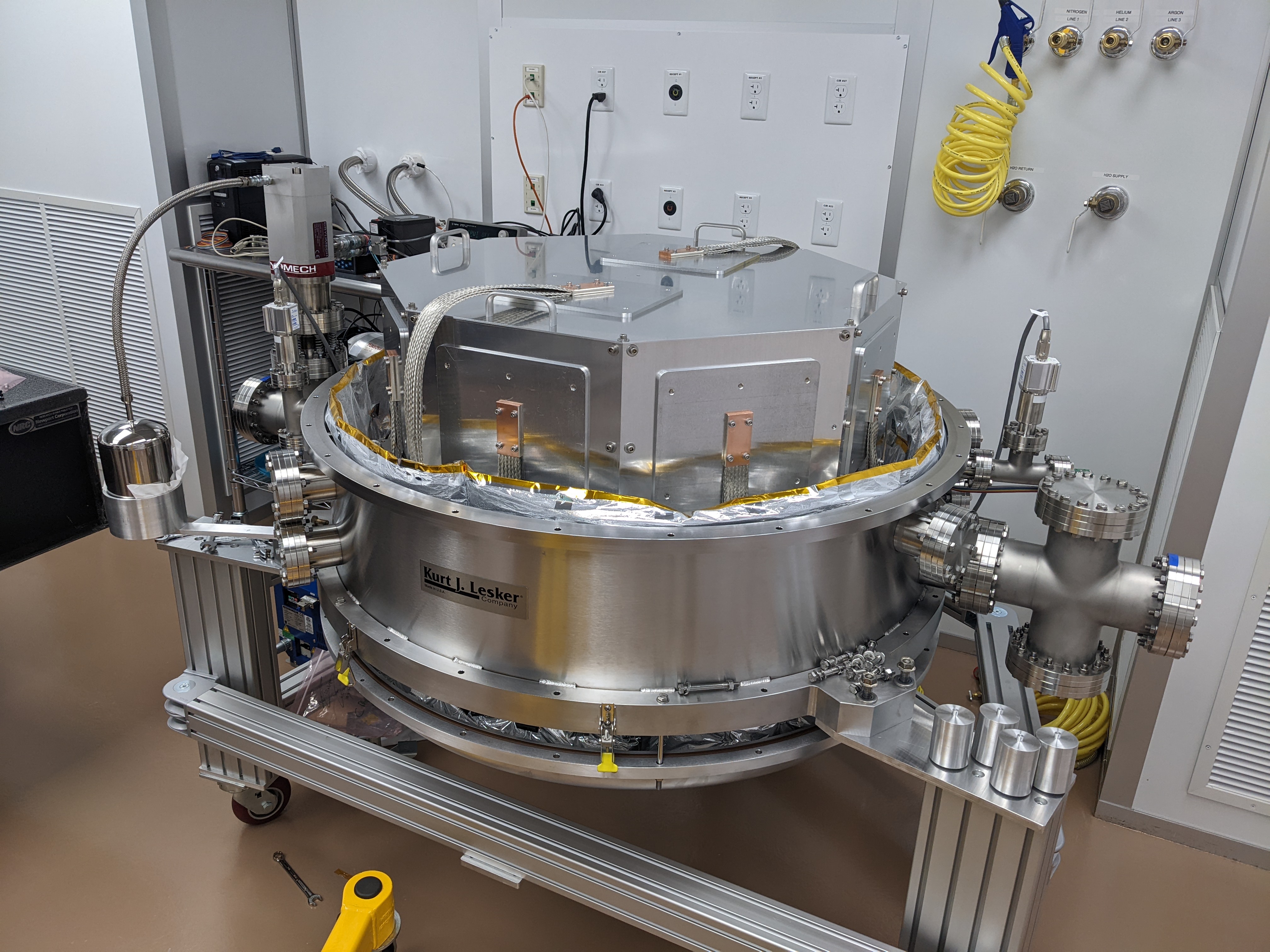}
   \end{tabular}
   \end{center}
   \caption[Integration of the cryostat] 
%>>>> use \label inside caption to get Fig. number with \ref{}
   { \label{fig:lab} 
The iLocater cryostat during system integration and performance testing.}
   \end{figure}

The iLocater cryostat functions to thermally, vibrationally, and physically isolate the spectrograph from its external environment to mitigate effects that would otherwise impact EPRV measurements. The system comprises a vacuum chamber with an internally mounted radiation shield that supports and surrounds the instrument spectrograph (Figure~\ref{fig:cryostat}).  The final design uses the same fundamental architecture as previous design iterations presented in \citenum{2016SPIE.9908E..73C} and has been optimized to house the completed spectrograph design. The assembled system is shown in Figure~\ref{fig:lab}.

The system vacuum chamber has been fabricated by Kurt J. Lesker Company and comprises three separate sections, each approximately 1m in diameter: a central stainless steel section is used to mechanically support the internal crystat components while upper and lower removable domes, fabricated from aluminum for handling considerations, complete the chamber. The central section of the chamber is designed to be vibrationally quiet and mechanically isolated from potential external sources of vibration present during operation. This minimizes any vibrations imparted onto the radiation shield and subsequently onto the spectrograph. The vacuum chamber is installed onto an extruded aluminum handling cart that attaches to the central section, allowing for system integration and transport. The cryostat will ultimately be installed on a vibrationally quiet mount attached to the telescope pier at the LBT.

All feedthroughs on the vacuum chamber are CF type. Those that are required for standard instrument operation (e.g. electrical, fiber, thermal) are housed within the central chamber section to minimize the need for de-cabling and similar processes during integration and instrument maintenance. Custom o-rings are used to make a vacuum tight seal between the chamber sections while still allowing efficient access as required. The inside of the vacuum chamber walls are lined with multi-layer insulation (MLI) (Section \ref{sec:MLI}) which reduces the thermal radiation loads on the internally mounted components. 

The instrument radiation shield is fabricated from aluminum and is actively controlled to provide a thermally stabilized environment for the spectrograph (see Section \ref{sec:thermal}). The aluminum alloy 3003 was chosen for the 8 wall panels and lid (0.1" thick) due to its improved thermal conductivity at cryogenic temperatures compared to the standard 6061 alloy \cite{NISTCryo}. The front three panels and lid are removable together as a single item to allow for efficient access to the spectrograph mounted inside, as shown in Figure~\ref{fig:cryostat}. The shield base plate is fabricated from 0.5” thick 6061 aluminum which is used to support the internally mounted spectrograph. While 3003 would have been preferred for thermal performance, its mechanical strength and limited form factors precluded its use for the base plate. 

All interfaces of the panels in the shield include grooved features to serve as a labyrinth ensuring a light tight fit. The internal face of each panel is painted with Z306 to improve the radiative coupling between the shield and the spectrograph while minimizing the impact of any stray light within the spectrograph system. The entire shield is mounted to the vacuum chamber wall using G10 standoffs for conduction isolation. These standoffs attach to interface tabs on the vacuum chamber wall and shield base plate, allowing for efficient installation and removal of the shield as needed.

\begin{figure} [ht]
   \begin{center}
   \begin{tabular}{c} %% tabular useful for creating an array of images 
   \includegraphics[width=\textwidth,trim={0 3cm 0 1.5cm},clip]{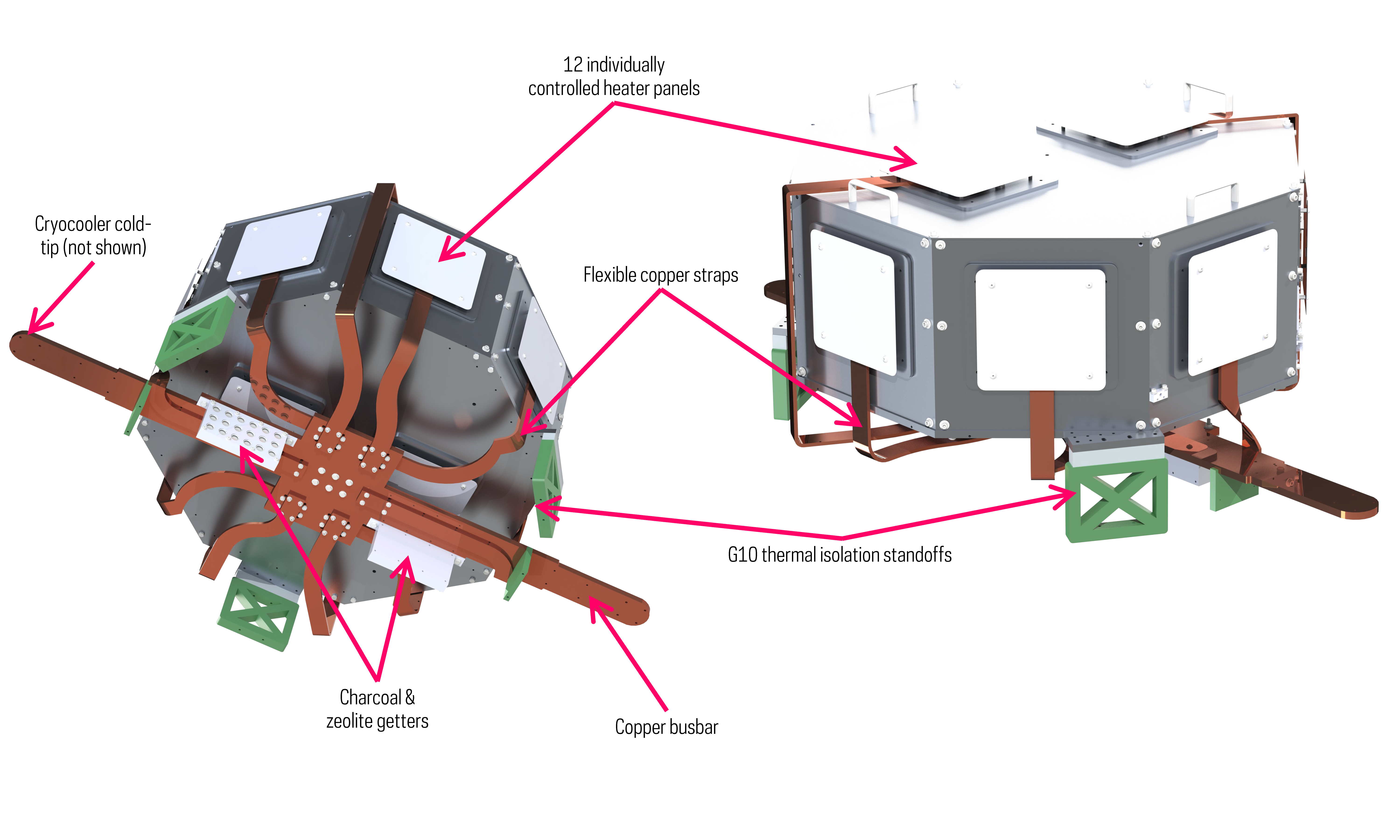}
   \end{tabular}
   \end{center}
   \caption[Cryostat thermal system] 
%>>>> use \label inside caption to get Fig. number with \ref{}
   { \label{fig:thermal} 
CAD rendering of the thermal system of the cryostat. A copper busbar connects the cold-tip of the PT90 cryocooler via a flexible copper braid interface (not shown). Flexible tin plated copper straps connect the busbar to the radiation shield.}
   \end{figure}

The cryostat thermal system (Figure~\ref{fig:thermal}) connects the cold-tip of a PT90 pulse-tube cryocooler from Cryomech to the radiation shield to provide system cooling. A pulse-tube cryocooler was selected for use due to its low vibration design. The cryocooler cold-head mounts vertically onto a side-port of the vacuum chamber via a spring-loaded bellow system and compact 6-way vacuum cube that provides access for the cold-head installation and removal. The inside of the cube and vacuum side of the cold-head are wrapped with MLI to reduce radiation loads. A flexible copper braided interface manufactured by Cryomech connects the cold-tip of the cryocooler to a copper ‘busbar’, minimizing vibration transmission while providing effective conduction. The busbar runs beneath the radiation shield and extends out through the vacuum chamber flange to the cold-tip. It is supported via a spring mechanism to minimize transmission of vibrations from the cryocooler onto the vacuum chamber wall. A total of 12 custom, flexible, tin-plated copper straps from Watteredge connect the busbar to 12 individual heater panels mounted on the shield wall, each one providing a direct conduction pathway to a thermal control point on the shield.

The vacuum pumping system for the chamber comprises a 594 L/min dry scroll pump (Ebara EV-A06) and 290 L/s  turbo pump (Leybold TURBOVAC 350ix). A pair of composite vacuum gauges (MKS 974B) mounted on the vacuum chamber provide redundant pressure readings on the chamber from $P=1\times10^{-8}-10^{3}\,\mathrm{Torr}$. The vacuum pumping hardware has been designed to achieve a pressure of $1\times10^{-3}\,\mathrm{Torr}$ pressure within 6 hours when accounting for outgassing from components within the system, particularly the MLI. A pair of getters are mounted onto the copper busbar to provide cryopumping capabilities and allowing the system to achieve an ultimate pressure of $P<5\times10^{-6}\,\mathrm{Torr}$.

\section{System Thermal Design \& Simulated Performance}
\label{sec:thermal}  % \label{} allows reference to this section

The thermal performance of the cryostat must enable a stable low temperature operating environment that meets the performance requirements of the spectrograph. The iLocater thermal system adopts a similar design and control strategy as employed by the Habitable-zone Planet Finder (HPF) and NEID instruments which have achieved sub-mK long-term stability \cite{2016ApJ...833..175S, 2019JATIS...5a5003R}. Stability is accomplished through a combination of radiative decoupling of the instrument and the vacuum chamber wall using MLI, with strong thermal pathways being provided by flexible copper straps between a cooling reservoir and the actively controlled radiation shield. iLocater’s busbar and pulse-tube cryocooler serve as its cooling reservoir, with the busbar providing thermal mass that acts as a low-pass filter for any temperature fluctuations of the cryocooler cold-tip. This is in comparison to the liquid nitrogen tank used in HPF and NEID, however, a cryocooler was required for iLocater due to the lower instrument operating temperature (Section \ref{sec:temperature}) and observatory operational constraints. Despite this change, similar levels of thermal stability to NEID and HPF are expected to be achieved with the iLocater system.

To achieve long-term stability, active thermal control is used at 12 control locations on the radiation shield (1 on each of the 8 wall panels, 2 on the lid and 2 on the base). Each location provides closed loop thermal control using identical hardware to the HPF and NEID thermal control systems. Two-wire (2N2222) and four-wire (CERNOX CX-1080) temperature sensors are mounted on the inside of the shield wall at each control location. The two-wire sensors provide precise relative measurements while the four-wire sensors provide absolute temperature values. Individual 3003 aluminum heater panels comprising wire-wound resistors are mounted on the corresponding outer panel face (Figure~\ref{fig:thermal}), and, together with custom electronics and commercial electronics (MicroK-250 from Isotech) as used in HPF and NEID, the system provides sub-mK temperature PID control at each location.

\subsection{Instrument Operating Temperature}
\label{sec:temperature}  % \label{} allows reference to this section

\begin{figure} [ht]
   \begin{center}
   \begin{tabular}{c} %% tabular useful for creating an array of images 
   \includegraphics[width=0.9\textwidth,trim={0 4.1cm 0 2cm},clip]{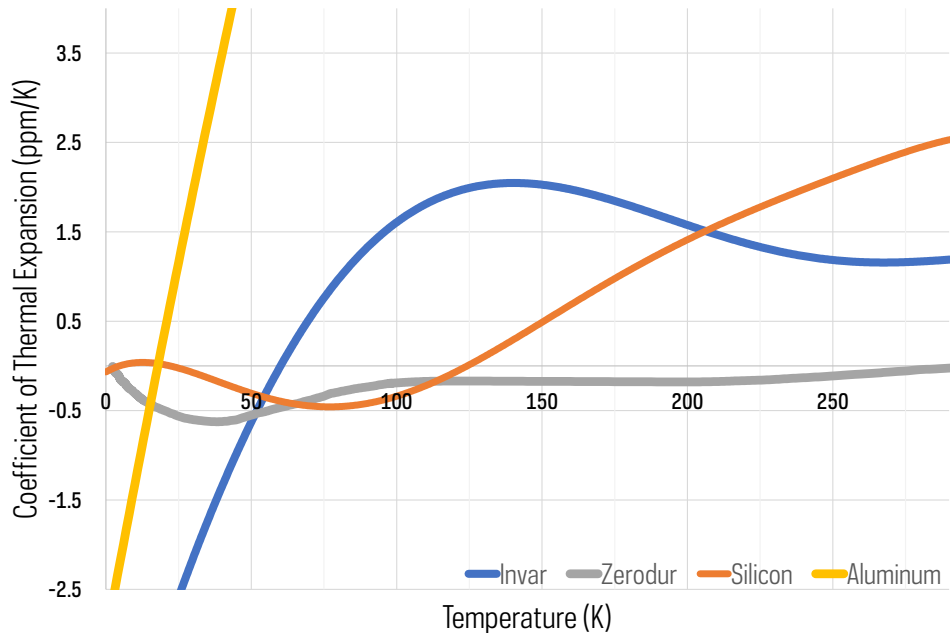}
   \end{tabular}
   \end{center}
   \caption[CTE of materials] 
%>>>> use \label inside caption to get Fig. number with \ref{}
   { \label{fig:cte} 
The coefficient of thermal expansion (CTE) of selected materials. The iLocater spectrograph optics are fabricated from silicon with its optomechanics being fabricated from Invar. By selecting an instrument operating temperature close to where the CTE is zero, the intrinsic stability of the optomechanical system can be maximized.}
   \end{figure}

The spectrograph operating temperature has been selected as $T = 80-100\,\mathrm{K}$ with the exact temperature being determined during full system integration. Temperature changes are achieved by adjusting the setpoint in the instrument thermal control system. The operating temperature range has been selected to both minimize the effects of thermal background on the detector while making the best use of the intrinsic thermal properties of the materials used in the optomechanical system. As the CTE of an individual material is a function of absolute temperature, by selecting an instrument operating temperature where the CTE of the materials used in the spectrograph are close to zero, the instrument gains improvements in stability (Figure~\ref{fig:cte}). For the Invar optomechanics and silicon optics and gratings used in the spectrograph, the $T=80-100\,\mathrm{K}$ temperature range is well optimized to minimize both global movements of optical components and intra-optics effects, such as line-space changes on the spectrograph gratings.

\subsection{Cryostat Thermal Model}
\label{sec:model}  % \label{} allows reference to this section

Extensive thermal modeling of the cryostat system has been undertaken, comprising both a simple lumped-element thermal circuit model and finite element analysis (FEA). Both models assume conservative estimates of performance. The thermal circuit model placed boundaries on overall thermal performance allowing optimization of key thermal system elements (Section \ref{sec:steadystate}). The FEA model has been compared to initial tests of the integrated system (Section \ref{sec:cooldowntest}) and demonstrates the as-built system will exceed simulated performance and achieve the required thermal environment. Both types of analysis have adopted the manufacturer nominal specification for the cooling capacity of the PT90 cryocooler used in the system. As this was only provided in the region of $T=30-80\,\mathrm{K}$, it was necessary to extrapolate the heat power curve up to 300K to assess the performance when cooling the system from room temperature.

\subsubsection{Steady State Thermal Circuit Model}
\label{sec:steadystate}  % \label{} allows reference to this section

To determine the appropriate conduction pathways needed in the cryostat system to achieve the instrument operating temperature, a detailed thermal circuit was constructed in Matlab. The model assumed a radiative heat load onto the radiation shield based upon the MLI performance of the HPF instrument and was designed to achieve a steady state temperature at approximately 10K below the instrument operating temperature \cite{2016ApJ...833..175S}. This approach allows a small amount of heat to be added to the system through the resistive heaters on the outside of the radiation shield to achieve thermal stability. The thermal model accounted for the temperature dependent cooling capacity of the cryocooler, conductive resistances between thermal interfaces in the conduction path, and varying thermal conductivity of copper at low temperatures. The model helped determine the minimum cross sectional area of the copper busbar and required cross-sectional area of the individual flexible straps to achieve desired cooling performance.

\subsubsection{Cooldown Thermal Analysis}
\label{sec:cooldownmodel}  % \label{} allows reference to this section

In order to assess and optimize the time required for the cryostat and spectrograph cooldown, a transient thermal FEA was performed using the simulation capabilities of SOLIDWORKS.  Analysis was completed in three phases: first the physical properties of the model were verified using a smaller test and development vacuum chamber. This housed a scaled down version of the internals of the cryostat and was constructed using the same materials and interface techniques as the finalized design.  The analysis of this system provided confidence in the assumptions (e.g. emissivity of the MLI and surfaces, thermal conductivity of components, thermal interface resistances, heat power curve of the cryocooler) used within the model. Second, the physical properties of the test chamber were incorporated into a thermal model of the final cryostat.  This model informed decisions such as radiation shield thickness, component mass, and determining if one cryocooler could provide sufficient cooling capacity for the system.  The model was also used as a reference during the cryostat manufacturing process to inform design decisions. This included the reduction of thermal mass in specific elements that were shown to drive the instrument cooldown time. The biggest change between the initial cooldown estimation model and the finalized system design is in the copper heat path between the cryocoolers and the flexible copper straps. Initially a heavier copper annulus component was selected due to system integration considerations, however, this led to an extended cooldown time. Through an optimization process driven by the thermal model, the annulus was eliminated in favor of a lighter interlocking copper bar (Figure~\ref{fig:thermal}) which achieved sufficient thermal performance and a reduced cooldown time. An approximate cooldown time for the spectrograph was 14 days, which brought the instrument to a stabilized final operating temperature of $T=85\,\mathrm{K}$. This was considered acceptable for performance and instrument stability.

\subsubsection{Thermal Perturbation Analysis}
\label{sec:perturbation}  % \label{} allows reference to this sectio

Additional analysis of the operating conditions of the cryostat was performed on the FEA model to better characterize the system response to transient events. These include a partial or total loss of power control, or changes in temperature of the vacuum chamber wall, with analyses focusing on the impact to instrument stability.  Transient thermal analysis was performed where the outer surface of one heater panel on the radiation shield was given a 1K temperature increase from the nominal operating temperature of the instrument (assumed to be 100K).  Results of this simulation indicate that the inner surface of the outer wall reached a steady state temperature of 101.2\,K after 1 minute of thermal perturbation, with the elevated temperature of the inner surface increasing the radiative heat load to the spectrograph.  Over the course of the perturbation study, the temperature change to the instrument was negligible, suggesting that minor adjustments in the temperature of the heaters does not affect the instrument temperature on short time scales.

\begin{figure} [ht]
   \begin{center}
   \begin{tabular}{c} %% tabular useful for creating an array of images 
   %\includesvg[width=\textwidth]{Figures/simulatedvsactualcooldown_edit_2.svg}
   \includegraphics[width=0.9\textwidth,trim={0 0 0 0},clip]{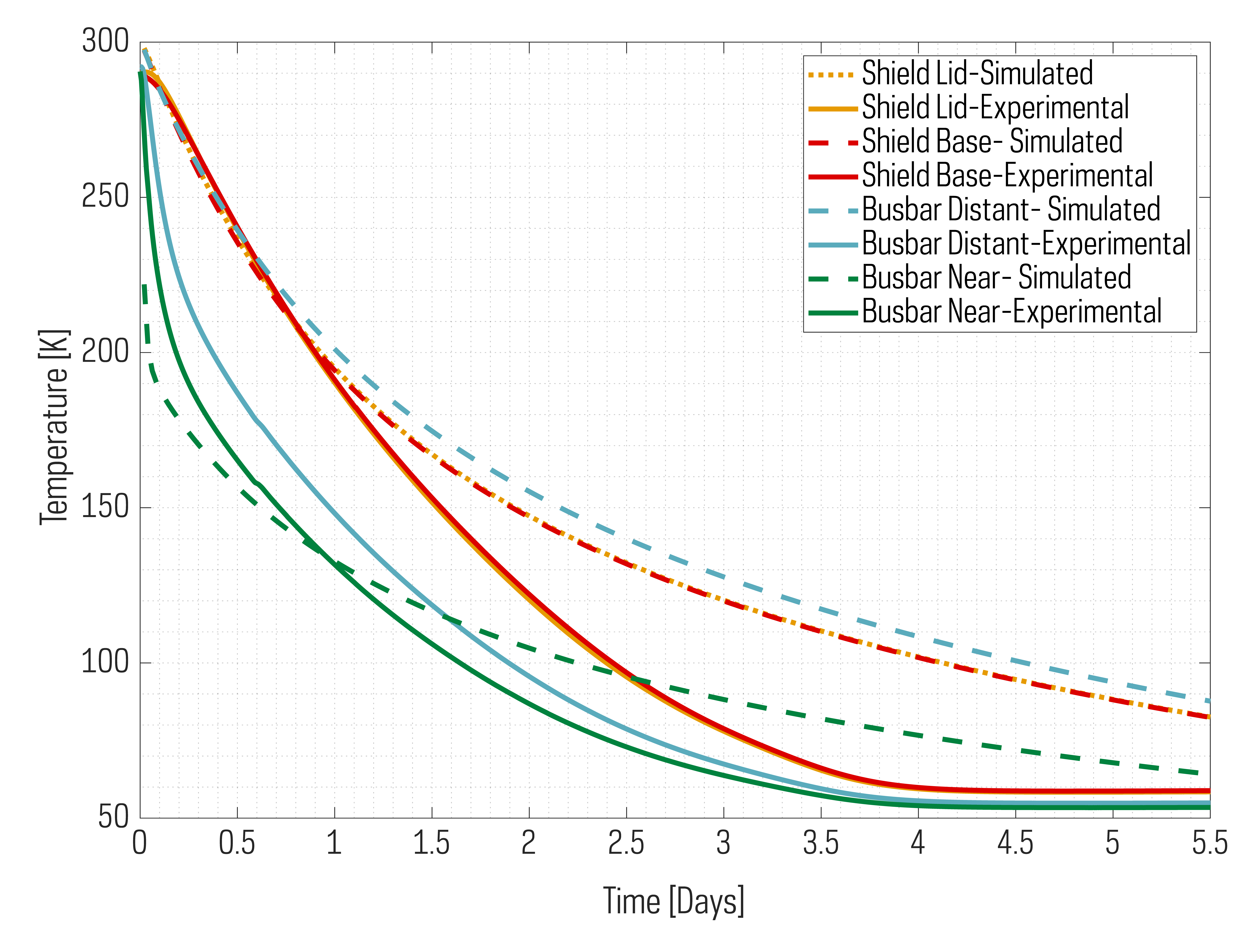}
   \end{tabular}
   \end{center}
   \caption[Initial system cooldown model] 
%>>>> use \label inside caption to get Fig. number with \ref{}
   { \label{fig:cooldown} 
The CAD model of the cryostat has been used to develop a simulated thermal model of performance in different system configurations. For the specific configuration shown, a test cool down of the cryostat was simulated without any spectrograph mass being present (dashed lines). The model assumes vendor specifications for cryocooler performance and conservative performance estimates for thermal interfaces and radiation loads. The measured cool down of the constructed system (solid lines) exceeds the model performance and demonstrates that iLocater will achieve its nominal operating temperature.}
   \end{figure} 

\subsection{MLI Design \& Fabrication}
\label{sec:MLI}  % \label{} allows reference to this section

The performance of the MLI blankets in iLocater is critical in reducing the radiative load from the vacuum chamber wall and enabling the instrument operating temperature to be reached. The blanket designs comprise five main sections that cover the entirety of the inner walls of the vacuum chamber. The chamber lids each have a circular blanket for the domed section and a cylindrical blanket that covers the vertical section. A separate cylindrical blanket is used for the central chamber section and is combined with smaller blankets that wrap other elements on the chamber walls, including the tabs that provide mechanical support of the internal components. A full-size mockup of the blankets was produced on vellum and installed on the chamber prior to the MLI blanket fabrication.

The blankets consist of alternating layers of aluminized mylar (14 layers) and nylon tulle (15 layers) and follow the same fabrication process used by HPF \cite{2016ApJ...833..175S}. Fourteen layers were included to ensure a sufficient reduction in radiative load from the chamber wall onto the shield. The nylon tulle acts as a spacer between the thin, highly reflective layers of mylar, providing durability to the blanket.  The layers are held together by plastic clothing tags which are inserted using a tag-gun and spaced a few inches apart around the edges of the blankets. Aluminum tape is added to the border of the blankets prior to tagging to enhance durability and minimize tears. The edges are then sealed using insulating Kapton tape. The MLI blankets are attached to the chamber using 1” non-shedding reclosable fasteners (3M SJ3560) secured both onto the blanket and, using bolts, onto standoffs welded onto the vacuum chamber wall. The individual blanket sections are secured to each other using stainless steel binder clips, allowing removal of the chamber domes and adjustment as needed. It is imperative that the adjacent sections of MLI are joined together with the same sides of each blanket touching each other to prevent a thermal short circuit that reduces the overall performance of the MLI. 

\subsection{Thermal Cool Down Testing}
\label{sec:cooldowntest}  % \label{} allows reference to this section

Initial laboratory cool down tests of the cryostat have been completed and assessed against the baseline of the thermal model performance. These tests did not include thermal mass inside the radiation shield, and instead focused on demonstrating that the instrument operating temperature can be reached using a single PT90 cryocooler. Figure~\ref{fig:cooldown} shows the results of the testing and a comparison to the FEA thermal model. Overall, the system without any spectrograph mass exceeded its simulated performance, achieving the instrument operating temperature in 2.5 days and a minimum temperature of 60K. While the addition of thermal mass to represent the spectrograph will increase the timeframe required to reach the operating temperature, this initial performance gives confidence that the cryostat can achieve its nominal temperature range.

\section{Conclusions}
\label{sec:conclusions}  % \label{} allows reference to this section

The iLocater cryostat is an optimized system that delivers the thermal environment needed to enable EPRV measurements. Through extensive simulations, design iterations and testing, the final system will deliver a stable cryogenic thermal environment and minimize paths for vibration transmission to the instrument spectrograph. The croystat system has already demonstrated it will achieve its operating temperature. 

The use of previously demonstrated thermal control hardware gives confidence that the system will achieve long-term sub-mK thermal stability, which is important to enable the key science goals of the iLocater spectrograph. Testing of the cryostat system is scheduled to continue over the next several months including the integration and optimization of the instrument spectrograph. This is in preparation for final delivery of the full iLocater system to the LBT in 2023.

\acknowledgments % equivalent to \section*{ACKNOWLEDGMENTS}       
 
This material is based upon work supported by the National Science Foundation under Grant No. 1654125 and 2108603. We thank Ted Wolfe for his vision and support of astronomy at Notre Dame. We thank the NEID and HPF teams, in particular Paul Robertson and Tyler Anderson, for their discussions and contributions regarding instrument thermal control.  

% References
\bibliography{report} % bibliography data in report.bib
\bibliographystyle{spiebib} % makes bibtex use spiebib.bst

\end{document}